\documentstyle[seceq,preprint]{ptptex}

\newcommand{\bra}[1]{\langle {#1} |}     
\newcommand{\ket}[1]{| {#1} \rangle}     
\newcommand{\wtilde}[1]{\widetilde{#1}} 
\newcommand{\lr}[1]{\langle{#1}\rangle} 

\title{
Deformed Boson Scheme including\\
Conventional $q$-Deformation in\\
Time-Dependent Variational Method. III
}
\subtitle{
Deformation of the su(2,1)-Algebra in Terms of\\
Three 
Kinds of Boson Operators}    

\author{
Atsushi {\sc Kuriyama}, 
Constan\c{c}a {\sc Provid\^encia}$^{*}$, \\
Jo\~ao da {\sc Provid\^encia}$^{*}$, Yasuhiko {\sc Tsue}$^{**}$ 
and Masatoshi {\sc Yamamura}
}

\inst{
Faculty of Engineering, Kansai University, Suita 564-8680, Japan\\
$^{*}$Departamento de Fisica, Universidade de Coimbra, P-3000 Coimbra, 
Portugal\\
$^{**}$Physics Division, Faculty of Science, Kochi University, 
Kochi 780-8520, Japan
}

\recdate{
}

\abst{
Basic idea presented in Parts (I) and (II) for the deformed boson 
scheme is applied to the case of the $su(2,1)$-algebra for 
describing many-body systems consisting of three kinds 
of boson operators. A possible form of the coherent state for 
the $su(2,1)$-algebra is generalized and the deformed $su(2,1)$-algebra 
is presented. As an application, the time-evolution of the 
statistically mixed state in a certain boson system, which has 
been already investigated by the present authors, is 
reinvestigated. 
}

\begin{document}

\maketitle

\section{Introduction}

In Part (I),\cite{KPPTY-I}
we developed a possible form of the deformed boson scheme 
which is suitable for treating many-body systems constituted of 
one kind of boson operator. The basic viewpoint can be found in the 
idea for a generalization of the boson coherent state. 
In this idea, a certain function of the boson number 
is introduced and we showed that the deformed boson scheme depends on 
the choice of the form of this function. 
In Part (II),\cite{KPPTY-II}
we investigated the deformed boson scheme for the system 
consisting of two kinds of boson operators. Naturally, the 
idea is a straightforward extension from that in (I). 
For the system in two kinds of boson operators, the 
$su(2)$- and the $su(1,1)$-algebra presented by Schwinger\cite{S} 
are well known. 
The algebras have been investigated by 
various authors in terms of the $q$-deformed algebras.\cite{Mac} 
We investigated this case by extending conventional boson 
coherent states to the forms which are suitable for the deformation 
of these algebras. 

Main aim of Part (III) is to formulate the deformation 
of the $su(2,1)$-algebra, the generators of which are 
expressed in terms of the bilinear forms for three kinds of boson 
operators. Naturally, the basic idea is the same as that adopted in (I) 
and (II). The present authors already formulated the $su(2,1)$-algebra 
in the above-mentioned case in Ref.\citen{KPTY2}, which is, 
hereafter, referred to as (A). 
Further, the present authors investigated a boson system 
interacting with an external harmonic oscillator as a possible 
description of statistically mixed state.\cite{KPTY3} 
Hereafter, Ref.\citen{KPTY3} is referred to as (B). 
The basic viewpoint adopted in (B) is found in the idea 
that the $su(2,1)$-algebra presented in (A) is deformed 
in the present sense. 
Then, it may be interesting to reinvestigate the deformation scheme 
given in (B) under the scheme extended from that adopted in (II). 

As was stressed in (II), our basic idea for the deformed  algebra is 
the use of the coherent state. In the case of the $su(2)$- and the 
$su(1,1)$-algebra, the coherent state is obtained in the form of 
exponential type of the superposition of the raising operator 
constructed on the state which vanishes under the operation 
of the lowering operator, i.e., the minimum weight state. 
In (III), we adopt the same idea as the above. However, there 
exists an essential difference 
between the forms in (II) and (III). 
In the case of the $su(2)$- and the $su(1,1)$-algebra, there exists 
only one which plays a role of the raising operator. 
However, in the case of the $su(2,1)$-algebra in the present form, 
there exist three operators which can be the candidate of the 
raising operator. But, for the construction of the coherent 
state, only two are necessary and, then, the coherent state 
depends on the choice of the two raising operators. 
In the present, we adopt the form which plays a central 
role in (B) and the deformation is performed for this 
coherent state. 
As an example of the application, the case discussed in (B) is 
reinvestigated and, as a natural consequence, we can describe the 
time-evolution of the statistically mixed state with the same results 
as those obtained in (B). 

In the next section, the $su(2,1)$-algebra in terms of three 
kinds of boson operators and its classical counterpart 
are recapitulated. Details have been given in (A). 
Section 3 is devoted to discussing the deformation of the 
coherent state, which plays a central role in \S 2. 
In \S 4, the $su(2,1)_q$-algebra is formulated in a certain 
special case which may be suitable for describing the 
time-evolution of the statistically mixed state of a boson 
system with an external harmonic oscillator. 
Finally, in \S 5, the formalism in (B) is reformulated 
in the framework of the present deformed boson scheme. 
Appendix is devoted to giving some mathematical formulae 
with the proof.

\section{The $su(2,1)$-algebra in terms of three kinds of boson operators 
and its classical counterpart}

As the preliminary for main parts in the present paper, 
first of all, we will recapitulate the $su(2,1)$-algebra 
in a special case : eight generators are expressed in terms of 
three kinds of boson operators, $({\hat a}, {\hat a}^*)$, 
$({\hat b}_1, {\hat b}_1^*)$ and $({\hat b}_2 , {\hat b}_2^*)$. 
Its details can be seen in (A), 
in which $({\hat b}_1,{\hat b}_1^*)$ and $({\hat b}_2, {\hat b}_2^*)$ 
should read $({\hat c},{\hat c}^*)$ and $({\hat d}, {\hat d}^*)$, 
respectively. The eight generators are expressed as bi-linear 
functions for the above boson operators in the following form : 
\begin{subequations}\label{2-1}
\begin{eqnarray}
& &{\hat T}_+^0 = {\hat a}^*{\hat b}_1^* \ , \qquad
{\hat T}_-^0 = {\hat b}_1{\hat a} \ , \qquad
{\hat T}_0 = ({\hat a}^*{\hat a}+{\hat b}_1{\hat b}_1^*)/2 \ , 
\label{2-1a}\\
& &{\hat S}_+^0 = {\hat b}_1^*{\hat b}_2 \ , \qquad
{\hat S}_-^0 = {\hat b}_2^*{\hat b}_1 \ , \qquad
{\hat S}_0 = ({\hat b}_1^*{\hat b}_1-{\hat b}_2^*{\hat b}_2)/2 \ , 
\label{2-1b}\\
& &{\hat R}_+ = {\hat a}^*{\hat b}_2^* \ , \qquad
{\hat R}_- = {\hat b}_2{\hat a} \ .
\label{2-1c}
\end{eqnarray}
\end{subequations}
The commutation relations for the above generators are given 
in the form 
\begin{subequations}\label{2-2}
\begin{eqnarray}
& &[{\hat T}_+^0 , {\hat T}_-^0 ]=- 2{\hat T}_0 \ , \qquad
[{\hat S}_+^0 , {\hat S}_-^0 ] = 2{\hat S}_0 \ , 
\label{2-2a}\\
& &[{\hat T}_{\pm}^0 , {\hat S}_{\mp}^0 ] = \mp {\hat R}_{\pm} \ , 
\qquad
[{\hat T}_{\pm}^0 , {\hat S}_{\pm}^0 ] = 0 \ , 
\label{2-2b}\\
& &[{\hat T}_0 , {\hat T}_\pm^0 ]= \pm {\hat T}_\pm^0 \ , \qquad
[{\hat S}_0 , {\hat T}_\pm^0 ] = \pm (1/2)\cdot{\hat T}_{\pm}^0 \ , 
\nonumber\\
& &[{\hat T}_0 , {\hat S}_{\pm}^0 ] = \pm (1/2)\cdot {\hat S}_{\pm}^0 \ , 
\qquad
[{\hat S}_0 , {\hat S}_{\pm}^0 ] = \pm {\hat S}_{\pm}^0 \ , 
\nonumber\\
& &[{\hat T}_0 , {\hat R}_{\pm} ] = \pm (1/2)\cdot {\hat R}_{\pm} \ , 
\qquad
[{\hat S}_0 , {\hat R}_{\pm} ] = \mp(1/2)\cdot {\hat R}_{\pm} \ , 
\nonumber\\
& &[{\hat T}_0 , {\hat S}_0 ] = 0 \ , 
\nonumber\\
& &[{\hat R}_+ , {\hat R}_- ] = -2 {\hat R}_0 \ , 
\qquad
(\ {\hat R}_0={\hat T}_0-{\hat S}_0 \ ) \ , 
\nonumber\\
& &[{\hat R}_\pm , {\hat T}_{\pm}^0 ] = 0 \ , 
\qquad
[{\hat R}_\pm , {\hat S}_{\pm}^0 ] = \mp {\hat T}_{\pm}^0 \ , 
\nonumber\\
& &[{\hat R}_\pm , {\hat T}_{\mp}^0 ] = \mp {\hat S}_{\mp}^0 \ , 
\qquad
[{\hat R}_\pm , {\hat S}_{\mp}^0 ] = 0 \ . 
\label{2-2c}
\end{eqnarray}
\end{subequations}
The Casimir operator ${\hat \Gamma}_{ab_1b_2}$, which commutes
with any generator, is expressed as 
\begin{eqnarray}
{\hat \Gamma}_{ab_1b_2}&=&
(1/2)\cdot ({\hat S}_0^2+{\hat T}_0^2+{\hat R}_0^2) \nonumber\\
& &+(3/4)[ ({\hat S}_-^0{\hat S}_+^0+{\hat S}_+^0{\hat S}_-^0)/2
-({\hat T}_-^0{\hat T}_+^0+{\hat T}_+^0{\hat T}_-^0)/2
-({\hat R}_-{\hat R}_+ +{\hat R}_+{\hat R}_-)/2] \nonumber\\
&=&{\hat K}^2-(3/2)\cdot {\hat K} \ , 
\label{2-3}\\
{\hat K}&=&({\hat b}_1{\hat b}_1^*+{\hat b}_2{\hat b}_2^*
-{\hat a}^*{\hat a})/2 \ .
\label{2-4}
\end{eqnarray}
The operator ${\hat K}$ commutes with any generator. 
We can see that $({\hat T}_\pm^0 , {\hat T}_0)$ and 
$({\hat R}_\pm , {\hat R}_0)$ form the $su(1,1)$-algebras, 
respectively, and $({\hat S}_\pm^0 , {\hat S}_0)$ the $su(2)$-algebra. 
The operators ${\hat T}$, ${\hat S}$ and ${\hat R}$, which are 
defined in the following, commute with 
$({\hat T}_\pm^0 , {\hat T}_0)$, $({\hat S}_\pm^0 , {\hat S}_0)$ 
and $({\hat R}_\pm, {\hat R}_0)$, respectively : 
\begin{eqnarray}\label{2-5}
& &{\hat T}=({\hat b}_1{\hat b}_1^*-{\hat a}^*{\hat a})/2 \ , 
\nonumber\\
& &{\hat S}=({\hat b}_1^*{\hat b}_1+{\hat b}_2^*{\hat b}_2)/2 \ , 
\nonumber\\
& &{\hat R}=({\hat b}_2{\hat b}_2^*-{\hat a}^*{\hat a})/2 \ . 
\end{eqnarray}
The operator ${\hat K}$ can be expressed in the form 
\begin{equation}\label{2-5-6}
{\hat K}=({\hat T}+{\hat S}+{\hat R}+1)/2 \ .
\end{equation}

As were done in (I) and (II), our starting task is to 
introduce a coherent state in the present system. 
In (A), we gave two forms for the coherent states. 
By modifying one of the two, we described, in (B), 
the statistically mixed state for a boson system interacting with an 
external harmonic oscillator. 
In the present notation, the form is expressed as follows : 
\begin{subequations}
\begin{eqnarray}
\ket{c^0}
&=&\left(\sqrt{\Gamma_0}\right)^{-1}
\exp\left(\gamma_1{\hat T}_+^0\right)
\exp\left(\gamma_2{\hat S}_+^0\right)
\exp\left(\delta{\hat b}_2^*\right)
\ket{0} \nonumber\\
&=&
\left(\sqrt{\Gamma_0}\right)^{-1}
\exp\left(\gamma_1{\hat a}^*{\hat b}_1^*\right)
\exp\left(\gamma_2{\hat b}_1^*{\hat b}_2\right)
\exp\left(\delta{\hat b}_2^*\right)
\ket{0} \ , 
\label{2-6}\\
\Gamma_0&=&
(1-|\gamma_1|^2)^{-1}
\exp\left((1-|\gamma_1|^2+|\gamma_2|^2)|\delta|^2\cdot 
(1-|\gamma_1|^2)^{-1}\right) \ .
\label{2-7}
\end{eqnarray}
\end{subequations}
Here, $\gamma_1$, $\gamma_2$ and $\delta$ denote complex parameters. 
In (A), we used $V$, $w$ and $v$, which are related with 
the present in the form 
$\gamma_1=V(\sqrt{1+|V|^2})^{-1}$, 
$\gamma_2=w(v\sqrt{1+|V|^2})^{-1}$ and $\delta=v$. 
The normalization constant (\ref{2-7}) can be rewritten 
in the form 
\begin{subequations}
\begin{eqnarray}
& &\Gamma_0=S(u, -v) e^{|\delta|^2} \ , \qquad
S(u, -v)=\frac{e^{v\frac{u}{1-u}}}{1-u} \ ,
\label{2-6b}\\
& &u=|\gamma_1|^2 \ , \qquad 
v=|\gamma_1|^{-2}|\gamma_2|^2|\delta|^2 \ .
\label{2-6c}
\end{eqnarray}
\end{subequations}
The function of $S(u, -v)$ is the generator of the 
Laguerre polynomials. 
The operators ${\hat T}_+^0$ and ${\hat S}_+^0$ 
play a role of generating the coherent state $\ket{c^0}$ on the 
state $\ket{m}=\exp(\delta{\hat b}_2^*)\ket{0}$, which obeys 
${\hat T}_-^0\ket{m}={\hat S}_-^0\ket{m}=0$. 
The state (\ref{2-6}) satisfies the following relations : 
\begin{subequations}\label{2-8}
\begin{eqnarray}
& &{\hat \gamma}_1^0\ket{c^0}=\gamma_1\ket{c^0} \ , 
\label{2-8a}\\
& &{\hat \gamma}_2^0\ket{c^0}=\gamma_2
(1-\epsilon({\hat N}_{b_1}+\epsilon)^{-1})\ket{c^0} \ , 
\label{2-9a}\\
& &{\hat \delta}^0\ket{c^0}=\delta\ket{c^0} \ . 
\label{2-10a}
\end{eqnarray}
\end{subequations}
Here, ${\hat \gamma}_1^0$, ${\hat \gamma}_2^0$ and 
${\hat \delta}^0$ are defined as 
\begin{subequations}\label{2-9}
\begin{eqnarray}
& &{\hat \gamma}_1^0=({\hat N}_{b_1}+1+\epsilon)^{-1}{\hat T}_-^0 \ ,
\label{2-8b}\\
& &{\hat \gamma}_2^0={\hat S}_-^0
({\hat N}_{b_2}+1+\epsilon)^{-1}\left(1-{\hat N}_a({\hat N}_{b_1}
+\epsilon)^{-1}\right) \ ,
\label{2-9b}\\
& &{\hat \delta}^0={\hat b}_2 \ , 
\label{2-10b}
\end{eqnarray}
\end{subequations}
\begin{equation}\label{2-11}
{\hat N}_a={\hat a}^*{\hat a} \ , \qquad
{\hat N}_{b_1}={\hat b}_1^*{\hat b}_1 \ , \qquad
{\hat N}_{b_2}={\hat b}_2^*{\hat b}_2 \ .
\end{equation}
The quantity $\epsilon$ is infinitesimal and 
$\epsilon({\hat N}_{b_1}+\epsilon)^{-1}$ denotes a projection 
operator, the role of which is explained in (II). 
In the same meaning as that given in (II), the complex parameters 
$\gamma_1$, $\gamma_2$ and $\delta$ are 
nothing but the eigenvalues of ${\hat \gamma}_1^0$, 
${\hat \gamma}_2^0$ and ${\hat \delta}^0$, respectively, 
and $\ket{c^0}$ is their eigenstate. 

By regarding $\ket{c^0}$ as a trial state, the present variational 
method starts with the following relation : 
\begin{equation}\label{2-12}
\delta\int\bra{c^0} i\partial_t - {\hat H} \ket{c^0} dt =0 \ .
\end{equation}
The state $\ket{c^0}$ satisfies the relation 
\begin{equation}\label{2-13}
\bra{c^0}i\partial_t\ket{c^0}
=(i/2)\cdot(x_1^*{\dot x}_1-{\dot x}_1^*x_1)
+(i/2)\cdot(x_2^*{\dot x}_2-{\dot x}_2^*x_2)
+(i/2)\cdot(y^*{\dot y}-{\dot y}^*y) \ .
\end{equation}
Here, $x_1$, $x_2$ and $y$ are defined as 
\begin{eqnarray}\label{2-14}
& &x_1=\gamma_1\sqrt{(\partial \Gamma_0/\partial |\gamma_1|^2)\cdot
\Gamma_0^{-1}} \ , 
\qquad 
(|x_1|^2=|\gamma_1|^2(\partial \Gamma_0/\partial |\gamma_1|^2)
\cdot \Gamma_0^{-1}) \nonumber\\
& &x_2=\gamma_2\sqrt{(\partial \Gamma_0/\partial |\gamma_2|^2)\cdot
\Gamma_0^{-1}} \ , 
\qquad 
(|x_2|^2=|\gamma_2|^2(\partial \Gamma_0/\partial |\gamma_2|^2)
\cdot \Gamma_0^{-1})\nonumber\\
& &y=\delta\sqrt{(\partial \Gamma_0/\partial |\delta|^2)\cdot
\Gamma_0^{-1}} \ .
\qquad 
(|y|^2=|\delta|^2(\partial \Gamma_0/\partial |\delta|^2)
\cdot \Gamma_0^{-1})
\end{eqnarray}
The relation (\ref{2-13}) tells us that $(x_1, x_1^*)$, 
$(x_2, x_2^*)$ and $(y, y^*)$ are boson-type canonical 
variables in classical mechanics. 
With the use of the relation (\ref{2-7}), we have 
\begin{eqnarray}\label{2-15}
& &x_1=\gamma_1\sqrt{1-|\gamma_1|^2+|\gamma_2|^2|\delta|^2}
\cdot (1-|\gamma_1|^2)^{-1} \ , \nonumber\\
& &x_2=\gamma_2|\delta|\cdot\left(
\sqrt{1-|\gamma_1|^2}\right)^{-1} \ , \nonumber\\
& &y=\delta\sqrt{1-|\gamma_1|^2+|\gamma_2|^2}
\cdot \left(\sqrt{1-|\gamma_1|^2}\right)^{-1} \ .
\end{eqnarray}
Inversely, the relation (\ref{2-15}) gives us 
\begin{eqnarray}\label{2-16}
& &\gamma_1=x_1\cdot \left(\sqrt{1+|x_1|^2+|x_2|^2}\right)^{-1} \ , 
\nonumber\\
& &\gamma_2=x_2\cdot \left(\sqrt{1+|x_1|^2+|x_2|^2}\right)^{-1}
\cdot \sqrt{1+|x_2|^2}\left(\sqrt{|y|^2-|x_2|^2}\right)^{-1} \ , 
\nonumber\\
& &\delta=y\cdot |y|^{-1}\sqrt{|y|^2-|x_2|^2} \ . 
\end{eqnarray}
In (A), we used the canonical variables $(X, X^*)$, 
$(Y, Y^*)$ and $(Z, Z^*)$, which are related to 
\begin{eqnarray}\label{2-17}
& &X=y\cdot |y|^{-1}\sqrt{|y|^2-|x_2|^2} \ (=v) \ , \nonumber\\
& &Y=x_1 \ (=V\sqrt{1+|w|^2}) \ , \nonumber\\
& &Z=x_2 y |y|^{-1} \ (=w) \ .
\end{eqnarray}

The expectation values of the generators for $\ket{c^0}$ 
are given in the following form : 
\begin{subequations}\label{2-18}
\begin{eqnarray}
& &\bra{c^0}{\hat T}_+^0\ket{c^0}
=x_1^*\sqrt{1+|x_1|^2+|x_2|^2} \ (\ =\langle T_+^0\rangle \ ) \ ,
\nonumber\\
& &\bra{c^0}{\hat T}_-^0\ket{c^0}
=x_1\sqrt{1+|x_1|^2+|x_2|^2} \ (\ =\langle T_-^0\rangle \ ) \ ,
\nonumber\\
& &\bra{c^0}{\hat T}_0\ket{c^0}
=(1+2|x_1|^2+|x_2|^2)/2 \ (\ =\langle T_0\rangle \ ) \ ,
\label{2-18a}\\
& &\bra{c^0}{\hat S}_+^0\ket{c^0}
=x_2^*\sqrt{|y|^2-|x_2|^2} 
\sqrt{(1+|x_1|^2+|x_2|^2)\cdot (1+|x_2|^2)^{-1}} 
\ (\ =\langle S_+^0\rangle \ ) \ ,
\nonumber\\
& &\bra{c^0}{\hat S}_-^0\ket{c^0}
=x_2\sqrt{|y|^2-|x_2|^2} 
\sqrt{(1+|x_1|^2+|x_2|^2)\cdot (1+|x_2|^2)^{-1}} 
\ (\ =\langle S_-^0\rangle \ ) \ ,
\nonumber\\
& &\bra{c^0}{\hat S}_0\ket{c^0}
=(|x_1|^2+2|x_2|^2-|y|^2)/2 \ (\ =\langle S_0\rangle \ ) \ ,
\label{2-18b}\\
& &\bra{c^0}{\hat R}_+\ket{c^0}
=x_1^*x_2\sqrt{|y|^2-|x_2|^2} \cdot
\left(\sqrt{1+|x_2|^2}\right)^{-1} 
\ (\ =\langle R_+\rangle \ ) \ ,
\nonumber\\
& &\bra{c^0}{\hat R}_-\ket{c^0}
=x_1x_2^*\sqrt{|y|^2-|x_2|^2} \cdot
\left(\sqrt{1+|x_2|^2}\right)^{-1} 
\ (\ =\langle R_-\rangle \ ) \ ,
\nonumber\\
& &\bra{c^0}{\hat R}_0\ket{c^0}
=(1+|x_1|^2-|x_2|^2+|y|^2)/2 \ (\ =\langle R_0\rangle \ ) \ .
\label{2-18c}
\end{eqnarray}
\end{subequations}
The relation between the Poisson brackets for the above 
expectation values and the commutation relations for the 
generators was discussed in (A), and, then, 
we omit the discussion. 
The expectation value of ${\hat K}$ defined in the relation 
(\ref{2-4}) is given by 
\begin{equation}\label{2-19}
\bra{c^0}{\hat K}\ket{c^0}=k=|y|^2/2+1 \ .
\end{equation}
We define the classical correspondence of the Casimir 
operator ${\hat \Gamma}_{ab_1b_2}$ defined in the relation (\ref{2-3}) 
in the form 
\begin{eqnarray}\label{2-20}
\Gamma_{ab_1b_2}&=&
(1/2)\cdot (\lr{S_0}^2+\lr{T_0}^2+\lr{R_0}^2) \nonumber\\
& &+(3/4)\cdot (\lr{S_+^0}\lr{S_-^0}
-\lr{T_+^0}\lr{T_-^0}-\lr{R_+}\lr{R_-} ) \ .
\end{eqnarray}
With the use of the relations (\ref{2-18}) and (\ref{2-19}), 
$\Gamma_{ab_1b_2}$ is written as 
\begin{equation}\label{2-21}
\Gamma_{ab_1b_2}=k^2-(3/2)\cdot k+3/4 \ .
\end{equation}
The above should be compared with the relation (\ref{2-3}). 
In this way, we obtain the classical counterpart of the 
$su(2,1)$-algebra expressed in terms of three kinds of boson 
operators.

Finally, we will sketch the relation between the set 
$({\hat \gamma}_1^0, {\hat \gamma}_2^0, {\hat \delta})$ and the 
set $(\gamma_1, \gamma_2, \delta)$. 
The commutation relations for the above operators are given, 
for example, in the form 
\begin{eqnarray}\label{2-22}
& &[{\hat \gamma}_1^0 , {\hat \gamma}_1^{0*}] 
=(\Delta_{{\hat N}_a}^{(+)}+\Delta_{{\hat N}_{b_1}}^{(+)}
+\Delta_{{\hat N}_a}^{(+)}\Delta_{{\hat N}_{b_1}}^{(+)}
)
({\hat \gamma}_1^{0*}{\hat \gamma}_1^0) \ , \nonumber\\
& &[{\hat \gamma}_2^0 , {\hat \gamma}_2^{0*}] 
=(\Delta_{{\hat N}_{b_1}}^{(+)}-\Delta_{{\hat N}_{b_2}}^{(-)}
-\Delta_{{\hat N}_{b_1}}^{(+)}\Delta_{{\hat N}_{b_2}}^{(-)})
({\hat \gamma}_2^{0*}{\hat \gamma}_2^0) \ , \nonumber\\
& &[{\hat \delta}^0 , {\hat \delta}^{0*}] 
=\Delta_{{\hat N}_{b_2}}^{(+)}
({\hat \delta}^{0*}{\hat \delta}^0) \ .
\end{eqnarray}
Here, $\Delta_{{\hat N}_c}^{(\pm)}$ for the operator 
${\hat N}_c={\hat c}^*{\hat c}$ $({\hat c}={\hat a},\ 
{\hat b}_1, \ {\hat b}_2)$ is defined as the difference for 
$F({\hat N}_c)$ : 
\begin{equation}\label{2-23}
\Delta_{{\hat N}_c}^{(\pm)}F({\hat N}_c)
=\pm \left[ F({\hat N}_c\pm 1)-F({\hat N}_c) \right] \ .
\end{equation}
The other combinations are omitted. The corresponding 
Poisson brackets are given as 
\begin{eqnarray}\label{2-24}
& &[\gamma_1 , \gamma_1^*]_P=(\partial_{N_a}+\partial_{N_{b_1}})
|\gamma_1|^2 \ , \nonumber\\
& &[\gamma_2 , \gamma_2^*]_P=(\partial_{N_{b_1}}-\partial_{N_{b_2}})
|\gamma_2|^2 \ , \nonumber\\
& &[\delta , \delta^*]_P=\partial_{N_{b_2}}
|\delta|^2 \ .
\end{eqnarray}
Here, the Poisson bracket is defined as 
\begin{eqnarray}\label{2-25}
[A, B]_P&=& 
(\partial_{x_1}A\cdot \partial_{x_1^*}B
-\partial_{x_1^*}A\cdot \partial_{x_1}B) \nonumber\\
& &+(\partial_{x_2}A\cdot \partial_{x_2^*}B
-\partial_{x_2^*}A\cdot \partial_{x_2}B) \nonumber\\
& &+(\partial_{y}A\cdot \partial_{y^*}B
-\partial_{y^*}A\cdot \partial_{y}B) \ .
\end{eqnarray}
The quantities $N_a$, $N_{b_1}$ and $N_{b_2}$ denote the 
expectation values of ${\hat N}_a$, ${\hat N}_{b_1}$ and 
${\hat N}_{b_2}$ for $\ket{c^0}$ : 
\begin{equation}\label{2-26}
N_a=|x_1|^2 \ , \qquad N_{b_1}=|x_1|^2+|x_2|^2 \ , \qquad
N_{b_2}=|y|^2-|x_2|^2 \ .
\end{equation}
Under the same argument as that given in (II), we can 
see that there exists the correspondence 
\begin{equation}\label{2-27}
{\hat \gamma}_1^0 \sim \gamma_1 \ , \qquad
{\hat \gamma}_2^0 \sim \gamma_2 \ , \qquad
{\hat \delta}^0 \sim \delta \ .
\end{equation}
This means that $\gamma_1$, $\gamma_2$ and $\delta$ 
introduced as the variational parameters are classical 
counterparts of the operators 
${\hat \gamma}_1^0$, ${\hat \gamma}_2^0$ and ${\hat \delta}$, 
respectively.

\section{Deformation of the state $\ket{c^0}$}

The most general form for the deformation of the state 
$\ket{c^0}$ given in the relation (\ref{2-6}) may be the 
following one : 
\begin{eqnarray}\label{3-1}
\ket{c}=\left(\sqrt{\Gamma}\right)^{-1}\!\!\!\!\!\!& &
\exp\left(\gamma_1{\hat a}^*{\wtilde f}({\hat N}_a)
\cdot {\hat b}_1^*{\wtilde g}_1({\hat N}_{b_1})\right) 
\nonumber\\
&\times&\exp\left(\gamma_2{\hat b}_1^*{\wtilde h}_1({\hat N}_{b_1})
\cdot {\wtilde g}_2({\hat N}_{b_2}){\hat b}_2\right) 
\nonumber\\
&\times&\exp\left(\delta{\hat b}_{2}^*{\wtilde h}_2({\hat N}_{b_2})^{-1}
\right)\ket{0}\ .
\end{eqnarray}
Properties of the function ${\wtilde f}(n)$, etc. are 
the similar to those presented in (II). 
The state $\ket{c}$ can be rewritten in the form 
\begin{eqnarray}
& &\ket{c}=\sqrt{\Gamma_0/\Gamma} \ \Omega({\hat N}_a, {\hat N}_{b_1}, 
{\hat N}_{b_2}) \ket{c^0} \ , 
\label{3-2}\\
& &\Omega({\hat N}_a, {\hat N}_{b_1}, {\hat N}_{b_2})
=f({\hat N}_a)g_1({\hat N}_{b_1})g_2({\hat N}_{b_2})^{-1}
\cdot h_1({\hat N}_{b_1}-{\hat N}_a)g_1
({\hat N}_{b_1}-{\hat N}_a)^{-1}
\nonumber\\
& &\qquad\qquad\qquad\qquad\qquad
\times
g_2({\hat N}_{b_1}+{\hat N}_{b_2}-{\hat N}_a)
h_2({\hat N}_{b_1}+{\hat N}_{b_2}-{\hat N}_a)^{-1} \ .
\end{eqnarray}
Here, $f(n)$, etc. is defined through the relation 
\begin{equation}\label{3-4}
{\wtilde f}(n)=f(n+1)f(n)^{-1}\ . \quad (n=0, 1, 2, \cdots)
\end{equation}
We described, in Ref.\citen{KPTY3}, the statistically 
mixed state for a boson system interacting with an external 
harmonic oscillator. In this papers, we will investigate another 
case : 
\begin{equation}\label{3-5}
h_1=g_1 \ , \qquad h_2=g_2 \ ,
\end{equation}
i.e., 
\begin{equation}\label{3-6}
\Omega({\hat N}_a, {\hat N}_{b_1}, {\hat N}_{b_2})
=f({\hat N}_a)g_1({\hat N}_{b_1})g_2({\hat N}_{b_2})^{-1}\ .
\end{equation}
Then, the state (\ref{3-1}) is reduced to 
\begin{eqnarray}\label{3-7}
\ket{c}=\left(\sqrt{\Gamma}\right)^{-1}\!\!\!\!\!& &
\exp\left(\gamma_1{\hat a}^*{\wtilde f}({\hat N}_a)
\cdot {\hat b}_1^*{\wtilde g}_1({\hat N}_{b_1})\right) 
\nonumber\\
&\times&\exp\left(\gamma_2\ \delta\ {\hat b}_1^*\ {\wtilde g}_1
({\hat N}_{b_1})\right) 
\nonumber\\
&\times&\exp\left(\delta\ {\hat b}_{2}^*\ {\wtilde g}_2
({\hat N}_{b_2})^{-1}
\right)\ket{0}\ .
\end{eqnarray}
Here, it should be noted that we used the relation 
\begin{equation}\label{3-8}
{\wtilde g}_2({\hat N}_{b_2}){\hat b}_2
\exp\left(\delta\ {\hat b}_2^*\ {\wtilde g}_2({\hat N}_{b_2})^{-1}
\right)\ket{0}
=\delta \exp\left(\delta\ {\hat b}_2^*\ {\wtilde g}_2({\hat N}_{b_2})^{-1}
\right)\ket{0} \ .
\end{equation}
The form (\ref{3-6}) is a direct extension from that used in 
(II) for the case of two kinds of boson operators. 
The state (\ref{3-7}) is expanded in the form 
\begin{eqnarray}\label{3-9}
\ket{c}&=&\left(\sqrt{\Gamma}\right)^{-1}
\sum_{n=0}^\infty \frac{f(n)\gamma_1^n}{n!} ({\hat a}^*)^n
\sum_{m=0}^\infty \frac{g_1(n+m)(\gamma_2\delta)^m}{m!} 
({\hat b}_1^*)^{(n+m)} \nonumber\\
& &\quad\qquad\qquad \times
\sum_{l=0}^\infty \frac{g_2(l)^{-1}\delta^l}{l!} 
({\hat b}_2^*)^l \ket{0} \ .
\end{eqnarray}
We impose the following condition : 
\begin{equation}\label{3-10}
f(0)=g_1(0)=g_2(0)=1 \ .
\end{equation}
The normalization constant $\Gamma$ is expressed as 
\begin{equation}\label{3-11}
\Gamma=\Gamma_{\gamma}\cdot 
\sum_{l=0}^{\infty}\frac{(|\delta|^2)^l}{l!} g_2(l)^{-2} \ .
\end{equation}
As is shown in the relation (\ref{A-31}), $\Gamma_{\gamma}$ is 
written as 
\begin{subequations}\label{3-11-2}
\begin{eqnarray}
& &\Gamma_\gamma = \sum_{r=0}^\infty 
\frac{G_{1r}(0)}{r!} u^r \sum_{n=0}^\infty \sum_{s=0}^n 
(-)^{n-s}\frac{F_n(0) n!}{(s!)^2(n-s)!} \nonumber\\
& &\qquad \times \left(\frac{\partial}{\partial u}\right)^{r+s}
\left[u^{n+s}S(u, -v)\right] \ , 
\label{3-11a}\\
& &f(n)^2=F(n)=\sum_{r=0}^\infty \frac{F_r(0)}{r!} n_r \ ,
\label{3-11b}\\
& &g_1(n)^2=G_1(n)=\sum_{r=0}^\infty \frac{G_{1r}(0)}{r!} n_r \ ,
\label{3-11c}\\
& &u=|\gamma_1|^2 \ , \qquad v=|\gamma_1|^{-2}|\gamma_2|^2|\delta|^2 \ .
\label{3-11d}
\end{eqnarray}
\end{subequations}
Here, $S(u, -v)$, $u$ and $v$ are defined in the relations 
(\ref{2-6b}) and (\ref{2-6c}), respectively. 
The meanings and the definitions of $G_{1r}(0)$, $F_r(0)$ 
and $n_r$ are given in Appendix.

Next, we define the operators ${\hat \gamma}_1$, ${\hat \gamma}_2$ 
and ${\hat \delta}$ in the form 
\begin{subequations}\label{3-12}
\begin{eqnarray}
& &{\hat \gamma}_1
=\Omega({\hat N}_a, {\hat N}_{b_1}, {\hat N}_{b_2}) \ 
{\hat \gamma}_1^0 \ \Omega({\hat N}_a, {\hat N}_{b_1}, 
{\hat N}_{b_2})^{-1} \ , 
\label{3-12a}\\
& &{\hat \gamma}_2
=\Omega({\hat N}_a, {\hat N}_{b_1}, {\hat N}_{b_2}) 
\ {\hat \gamma}_2^0\ \Omega({\hat N}_a, {\hat N}_{b_1}, 
{\hat N}_{b_2})^{-1} \ , 
\label{3-13a}\\
& &{\hat \delta}
=\Omega({\hat N}_a, {\hat N}_{b_1}, {\hat N}_{b_2}) 
\ {\hat \delta}^0\ \Omega({\hat N}_a, {\hat N}_{b_1}, 
{\hat N}_{b_2})^{-1} \ . 
\label{3-14a}
\end{eqnarray}
\end{subequations}
Here, ${\hat \gamma}_1^0$, ${\hat \gamma}_2^0$ and ${\hat \delta}^0$ 
are given in the relations (\ref{2-8b}), (\ref{2-9b}) and 
(\ref{2-10b}), respectively. Since we have the relations 
(\ref{2-8a}), (\ref{2-9a}) and (\ref{2-10a}), the following 
equations are derived : 
\begin{subequations}
\begin{eqnarray}
& &{\hat \gamma}_1\ket{c}=\gamma_1 \ket{c} \ , 
\label{3-12b}\\
& &{\hat \gamma}_2\ket{c}=\gamma_2(1-\epsilon({\hat N}_{b_1}
+\epsilon)^{-1}) \ket{c} \ , 
\label{3-13b}\\
& &{\hat \delta}\ket{c}=\delta \ket{c} \ . 
\label{3-14b}
\end{eqnarray}
\end{subequations}
The commutation relations for the above operators are given, 
for example, in the form 
\begin{eqnarray}\label{3-15}
& &[{\hat \gamma}_1 , {\hat \gamma}_1^* ]
=\left(\Delta_{{\hat N}_a}^{(+)}+\Delta_{{\hat N}_{b_1}}^{(+)}
+\Delta_{{\hat N}_a}^{(+)}\Delta_{{\hat N}_{b_1}}^{(+)}\right)
({\hat \gamma}_1^*{\hat \gamma}_1) \ , \nonumber\\
& &[{\hat \gamma}_2 , {\hat \gamma}_2^* ]
=\left(\Delta_{{\hat N}_{b_1}}^{(+)}-\Delta_{{\hat N}_{b_2}}^{(-)}
-\Delta_{{\hat N}_{b_1}}^{(+)}\Delta_{{\hat N}_{b_2}}^{(-)}\right)
({\hat \gamma}_2^*{\hat \gamma}_2) \ , \nonumber\\
& &[{\hat \delta} , {\hat \delta}^* ]
=\Delta_{{\hat N}_{b_2}}^{(+)}
({\hat \delta}^*{\hat \delta}) \ .
\end{eqnarray}
The relation (\ref{3-15}) is completely of the same form as 
that shown in the relation (\ref{2-22}). 

We can prove that $\ket{c}$ also satisfies the relation 
\begin{equation}\label{3-16}
\bra{c}i\partial_t \ket{c}
=(i/2)(x_1^*{\dot x}_1-{\dot x}_1^* x_1)
+(i/2)(x_2^*{\dot x}_2-{\dot x}_2^* x_2)
+(i/2)(y^*{\dot y}-{\dot y}^* y) \ .
\end{equation}
Here, $x_1$, $x_2$ and $y$ are defined as 
\begin{subequations}\label{3-17}
\begin{eqnarray}
& &x_1=\gamma_1 \sqrt{(\partial \Gamma/\partial |\gamma_1|^2)
\cdot \Gamma^{-1}} \ , \qquad
(|x_1|^2=|\gamma_1|^2\cdot(\partial \Gamma/\partial |\gamma_1|^2)
\cdot\Gamma^{-1}) \quad 
\label{3-17a}\\
& &x_2=\gamma_2 \sqrt{(\partial \Gamma/\partial |\gamma_2|^2)
\cdot \Gamma^{-1}} \ , \qquad
(|x_2|^2=|\gamma_2|^2\cdot(\partial \Gamma/\partial |\gamma_2|^2)
\cdot\Gamma^{-1}) \quad
\label{3-17b}\\
& &y=\delta \sqrt{(\partial \Gamma/\partial |\delta|^2)
\cdot \Gamma^{-1}} \ . \qquad\qquad
(|y|^2=|\delta|^2\cdot(\partial \Gamma/\partial |\delta|^2)
\cdot\Gamma^{-1}) 
\label{3-17c}
\end{eqnarray}
\end{subequations}
The relation (\ref{3-16}) tells us that 
$(x_1, x_1^*)$, $(x_2, x_2^*)$ and $(y, y^*)$ are boson-type 
canonical variables. 
In the case of $\ket{c^0}$, we can express $\gamma_1$, $\gamma_2$ and 
$\delta$ as functions of $x_1$, $x_2$ and $y$, which are shown 
in the relation (\ref{2-16}). 
However, in the present case, it is in general impossible. 
The expectation values of ${\hat N}_a$, ${\hat N}_{b_1}$ and 
${\hat N}_{b_2}$ for $\ket{c}$, which we denote 
$N_a$, $N_{b_1}$ and $N_{b_2}$, are given as 
\begin{eqnarray}\label{3-18}
& &N_a=|\gamma_1|^2\frac{\partial \Gamma}{\partial |\gamma_1|^2}
\cdot \Gamma^{-1}=|x_1|^2 \ , \nonumber\\
& &N_{b_1}=|\gamma_1|^2\frac{\partial \Gamma}{\partial |\gamma_1|^2}
\cdot \Gamma^{-1}+|\gamma_2|^2\frac{\partial \Gamma}{
\partial |\gamma_2|^2}\cdot \Gamma^{-1}=|x_1|^2+|x_2|^2 \ , \nonumber\\
& &N_{b_2}=|\delta|^2\frac{\partial \Gamma}{\partial |\delta|^2}
\cdot \Gamma^{-1}-|\gamma_2|^2\frac{\partial \Gamma}{
\partial |\gamma_2|^2}\cdot \Gamma^{-1}=|y|^2-|x_2|^2 \ .
\end{eqnarray}
The form (\ref{3-18}) is completely the same as that shown 
in the relation (\ref{2-26}) for $\ket{c^0}$. 
Therefore, in the same logic adopted in (II), we can conclude that 
${\hat \gamma}_1$, ${\hat \gamma}_2$ and ${\hat \delta}$ correspond 
to $\gamma_1$, $\gamma_2$ and $\delta$, respectively. 

We can calculate the expectation values of various operators 
for $\ket{c}$, the examples of which are shown in the relation 
(\ref{3-18}). 
The other examples are the expectation values of the operators 
${\hat \sigma}_+=
{\hat b}_1^*\ {\wtilde g}_1({\hat N}_{b_1})\cdot {\wtilde g}_2
({\hat N}_{b_2}){\hat b}_2$ and 
${\hat \tau}_-={\wtilde g}_1({\hat N}_{b_1})^{-1}{\hat b}_1\cdot
{\wtilde f}({\hat N}_a)^{-1}{\hat a}$. 
The straightforward calculation gives us 
\begin{subequations}\label{3-19}
\begin{eqnarray}
& &{\hat \sigma}_+\ket{c}=\gamma_2^*\frac{1}{|\gamma_2|^2}
({\hat N}_{b_1}-{\hat N}_{a})\ket{c} \ , 
\label{3-19a}\\
& &{\hat \tau}_-\ket{c}=\gamma_1
({\hat N}_{b_1}+1)\ket{c} \ . 
\label{3-19b}
\end{eqnarray}
\end{subequations}
Then, with the use of the relation (\ref{3-18}), we have 
\begin{subequations}\label{3-20}
\begin{eqnarray}
& &\bra{c}{\hat \sigma}_+\ket{c}
=x_2^*\sqrt{\frac{\partial\Gamma}{\partial |\gamma_2|^2}
\cdot \Gamma^{-1}} \ , 
\label{3-20a}\\
& &\bra{c}{\hat \tau}_-\ket{c}
=x_1\sqrt{\left(\frac{\partial\Gamma}{\partial |\gamma_1|^2}\right)^{-1}
\cdot \Gamma}\cdot(1+|x_1|^2+|x_2|^2) \ . 
\label{3-20b}
\end{eqnarray}
\end{subequations}
The operator ${\hat \sigma}_+$ 
is nothing but 
${\hat S}_+$ introduced in the next section and 
in \S 5, we investigate the form (\ref{3-20a}). 
%
%

\section{The $su(2,1)_q$-algebra in the present deformed boson 
scheme}

It may be an interesting problem to investigate 
the $su(2,1)_q$-algebra in the present deformed boson scheme. 
If our understanding is correct, the algebra 
such as the $su(2,1)_q$-algebra has not been investigated 
by anyone. Our starting idea for formulating the $su(2,1)_q$-algebra 
is as follows : 
As was shown in the previous section, the state $\ket{c}$ is 
deformed from $\ket{c^0}$ through the three parts. 
As is clear from the state $\ket{c^0}$ shown in the relation 
(\ref{2-6}), the parts ${\hat T}_+^0$, ${\hat S}_+^0$ and 
$\ket{m}=\exp(\delta {\hat b}_2^*)\ket{0}$ are deformed. 
Therefore, for the deformation of the generators 
of the $su(2,1)$-algebra, ${\hat T}_{\pm}^0$ and ${\hat S}_{\pm}^0$ 
are deformed to ${\hat T}_{\pm}$ and ${\hat S}_{\pm}$ from 
the outside and the remaining generators $2{\hat T}_0$, 
$2{\hat S}_0$ and ${\hat R}_{\pm}$ are deformed through the 
commutation relation 
\begin{subequations}\label{4-1}
\begin{eqnarray}
& &[{\hat T}_+ , {\hat T}_- ] = - [2{\hat T}_0]_q \ , \qquad
[{\hat S}_+ , {\hat S}_- ] = + [2{\hat S}_0]_q \ , 
\label{4-1a}\\
& &[{\hat T}_{\pm} , {\hat S}_{\mp} ] = \mp [{\hat R}_\pm]_q \ . 
\label{4-1b}
\end{eqnarray}
\end{subequations}
The above is our starting idea for the deformation. 

Following the basic form shown in the relation 
(II$\cdot$4$\cdot$3), we define the operators
\begin{eqnarray}\label{4-2}
& &{\hat \alpha}=\Omega({\hat N}_a, {\hat N}_{b_1}, {\hat N}_{b_2})^{-1} 
\ {\hat a}\ 
\Omega({\hat N}_a, {\hat N}_{b_1}, {\hat N}_{b_2}) \ , \nonumber\\
& &{\hat \beta}_1=\Omega({\hat N}_a, {\hat N}_{b_1}, {\hat N}_{b_2})^{-1} 
\ {\hat b}_1\ 
\Omega({\hat N}_a, {\hat N}_{b_1}, {\hat N}_{b_2}) \ , \nonumber\\
& &{\hat \beta}_2
=\Omega({\hat N}_a, {\hat N}_{b_1}, {\hat N}_{b_2}) 
\ {\hat b}_2\ 
\Omega({\hat N}_a, {\hat N}_{b_1}, {\hat N}_{b_2})^{-1} \ . 
\end{eqnarray}
Here, $\Omega({\hat N}_a, {\hat N}_{b_1}, {\hat N}_{b_2})$ is 
given in the relation (\ref{3-6}). 
Then, for ${\hat T}_{\pm}$ and ${\hat S}_{\pm}$, we give the forms 
\begin{subequations}\label{4-3}
\begin{eqnarray}
& &{\hat T}_+={\hat \alpha}^*{\hat \beta}_1^* \ , \qquad
{\hat T}_-={\hat \beta}_1{\hat \alpha} \ ,
\label{4-3a}\\
& &{\hat S}_+={\hat \beta}_1^*{\hat \beta}_2 \ , \qquad
{\hat S}_-={\hat \beta}_2^*{\hat \beta}_1 \ .
\label{4-3b}
\end{eqnarray}
\end{subequations}
The explicit forms of ${\hat T}_{\pm}$ and ${\hat S}_{\pm}$ 
defined in the above are written as 
\begin{subequations}\label{4-4}
\begin{eqnarray}
& &{\hat T}_+=f({\hat N}_a) g_1({\hat N}_{b_1})
\ {\hat T}_+^0\ f({\hat N}_a)^{-1}g_1({\hat N}_{b_1})^{-1} \ , \qquad
{\hat T}_+^0={\hat a}^*{\hat b}_1^* \ , \nonumber\\
& &{\hat T}_-=f({\hat N}_a)^{-1}g_1({\hat N}_{b_1})^{-1}
\ {\hat T}_-^0\ f({\hat N}_a)g_1({\hat N}_{b_1}) \ , \qquad
{\hat T}_-^0={\hat b}_1{\hat a} \ , 
\label{4-4a}\\
& &{\hat S}_+=g_1({\hat N}_{b_1})g_2({\hat N}_{b_2})^{-1}
\ {\hat S}_+^0\ g_1({\hat N}_{b_1})^{-1}g_2({\hat N}_{b_2}) \ , \qquad
{\hat S}_+^0={\hat b}_1^*{\hat b}_2 \ , \nonumber\\
& &{\hat S}_-=g_1({\hat N}_{b_1})^{-1}g_2({\hat N}_{b_2})
\ {\hat S}_-^0\  g_1({\hat N}_{b_1})g_2({\hat N}_{b_2})^{-1} \ , \qquad
{\hat S}_-^0={\hat b}_2^*{\hat b}_1 \ . 
\label{4-4b}
\end{eqnarray}
\end{subequations}
With the use of the relations (\ref{4-4a}) and (\ref{4-4b}), 
$[2{\hat T}_0]_q$, $[2{\hat S}_0]_q$ and $[{\hat R}_{\pm}]_q$ 
are obtained by the commutation relations (\ref{4-1a}) and 
(\ref{4-1b}) :
\begin{subequations}\label{4-5}
\begin{eqnarray}
& &[2{\hat T}_0]_q=[{\hat N}_a+1]_f [{\hat N}_{b_1}+1]_{g_1}
-[{\hat N}_a]_f [{\hat N}_{b_1}]_{g_1} \ , 
\label{4-5a}\\
& &[2{\hat S}_0]_q=[{\hat N}_{b_1}]_{g_1} [{\hat N}_{b_2}+1]_{g_2}
-[{\hat N}_{b_1}+1]_{g_1} [{\hat N}_{b_2}]_{g_2} \ , 
\label{4-5b}
\end{eqnarray}
\end{subequations}
\begin{eqnarray}\label{4-6}
& &[{\hat R}_+]_q={\hat \alpha}^*{\hat \beta}_2^* \left[
[{\hat N}_{b_1}+1]_{g_1}-[{\hat N}_{b_1}]_{g_1}\right] \ , \qquad
\quad\nonumber\\
& &[{\hat R}_-]_q={\hat \beta}_2{\hat \alpha} \left[
[{\hat N}_{b_1}+1]_{g_1}-[{\hat N}_{b_1}]_{g_1}\right] \ . 
\end{eqnarray}
Here, $[x]_f$, $[x]_{g_1}$ and $[x]_{g_2}$ are defined by 
\begin{eqnarray}\label{4-7}
& &[x]_f=x f(x)^{2} f(x-1)^{-2} \ , \nonumber\\
& &[x]_{g_1}=x g_1(x)^{2} g_1(x-1)^{-2} \ , \qquad
[x]_{g_2}=x g_2(x)^{2} g_2(x-1)^{-2} \ .
\end{eqnarray}
The operators ${\hat \alpha}$ and ${\hat \beta}_2$ are defined 
in the relation (\ref{4-2}) : 
\begin{equation}\label{4-8}
{\hat \alpha}=f({\hat N}_a+1) f({\hat N}_a)^{-1}\ {\hat a} \ , 
\qquad
{\hat \beta}_2=g_2({\hat N}_{b_2}+1) 
g_2({\hat N}_{b_2})^{-1}\ {\hat b}_2 \ . 
\end{equation}

As was done in (II), let us introduce the operator 
$({\hat E}_c , {\hat E}_c^*)$ for $c=a$, $b_{1}$ and $b_2$ 
defined as 
\begin{equation}\label{4-9}
{\hat E}_c = \left(\sqrt{{\hat N}_c +1}\right)^{-1}{\hat c} \ ,
\qquad
{\hat E}_c^* = {\hat c}^*\left(\sqrt{{\hat N}_c +1}\right)^{-1} \ . 
\qquad ({\hat N}_c={\hat c}^*{\hat c})
\end{equation} 
The property is as follows : 
\begin{equation}\label{4-10}
{\hat E}_c {\hat E}_c^* = 1 \ , \qquad
{\hat N}_c {\hat E}_c^*{\hat E}_c 
={\hat E}_c^*{\hat E}_c{\hat N}_c = {\hat N}_c \ .
\end{equation}
With the use of the operator $({\hat E}_c , {\hat E}_c^*)$, 
the deformed generators in the present algebra can be expressed as 
\begin{subequations}\label{4-11}
\begin{eqnarray}
& &{\hat T}_+={\hat E}_a^*{\hat E}_{b_1}^*
\sqrt{[{\hat T}_0-{\hat T}+1]_f [{\hat T}_0+{\hat T}]_{g_1}}
=\sqrt{[{\hat T}_0-{\hat T}]_f [{\hat T}_0+{\hat T}-1]_{g_1}}
{\hat E}_a^*{\hat E}_{b_1}^* \ , \nonumber\\
& &{\hat T}_-={\hat E}_{b_1}{\hat E}_{a}
\sqrt{[{\hat T}_0-{\hat T}]_f [{\hat T}_0+{\hat T}-1]_{g_1}}
=\sqrt{[{\hat T}_0-{\hat T}+1]_f [{\hat T}_0+{\hat T}]_{g_1}}
{\hat E}_{b_1}{\hat E}_{a} \ , \nonumber\\
& &[2{\hat T}_0]_q=[{\hat T}_0-{\hat T}+1]_f [{\hat T}_0+{\hat T}]_{g_1}
-[{\hat T}_0-{\hat T}]_f [{\hat T}_0+{\hat T}-1]_{g_1} \ , 
\label{4-11a}\\
& &{\hat S}_+={\hat E}_{b_1}^*{\hat E}_{b_2}
\sqrt{[{\hat S}+{\hat S}_0+1]_{g_1} [{\hat S}-{\hat S}_0]_{g_2}}
=\sqrt{[{\hat S}+{\hat S}_0]_{g_1} [{\hat S}-{\hat S}_0+1]_{g_2}}
{\hat E}_{b_1}^*{\hat E}_{b_2} \ , \nonumber\\
& &{\hat S}_-={\hat E}_{b_2}^*{\hat E}_{b_1}
\sqrt{[{\hat S}+{\hat S}_0]_{g_1} [{\hat S}-{\hat S}_0+1]_{g_2}}
=\sqrt{[{\hat S}+{\hat S}_0+1]_{g_1} [{\hat S}-{\hat S}_0]_{g_2}}
{\hat E}_{b_2}^*{\hat E}_{b_1} \ , \nonumber\\
& &[2{\hat S}_0]_q=
[{\hat S}+{\hat S}_0]_{g_1} [{\hat S}-{\hat S}_0+1]_{g_2}
-[{\hat S}+{\hat S}_0+1]_{g_1} [{\hat S}-{\hat S}_0]_{g_2}\ , 
\label{4-11b}\\
& &[{\hat R}_+]_q={\hat E}_{a}^*{\hat E}_{b_2}^*
\sqrt{[{\hat R}_0-{\hat R}+1]_{f} [{\hat R}_0+{\hat R}]_{g_2}}
\left[ [{\hat S}+{\hat S}_0+1]_{g_1}-[{\hat S}+{\hat S}_0]_{g_1}\right] 
\nonumber\\
& &\qquad\quad
=\sqrt{[{\hat R}_0-{\hat R}]_{f} [{\hat R}_0+{\hat R}-1]_{g_2}}
\left[ [{\hat S}+{\hat S}_0+1]_{g_1}-[{\hat S}+{\hat S}_0]_{g_1}
\right] 
{\hat E}_{a}^*{\hat E}_{b_2}^* \ , \nonumber\\
& &[{\hat R}_-]_q={\hat E}_{b_2}{\hat E}_{a}
\sqrt{[{\hat R}_0-{\hat R}]_{f} [{\hat R}_0+{\hat R}-1]_{g_2}}
\left[ [{\hat S}+{\hat S}_0+1]_{g_1}-[{\hat S}+{\hat S}_0]_{g_1}\right] 
\nonumber\\
& &\qquad\quad
=\sqrt{[{\hat R}_0-{\hat R}+1]_{f} [{\hat R}_0+{\hat R}]_{g_2}}
\left[ [{\hat S}+{\hat S}_0+1]_{g_1}-[{\hat S}+{\hat S}_0]_{g_1}
\right] 
{\hat E}_{b_2}{\hat E}_{a} \ . \qquad\quad
\label{4-11c}
\end{eqnarray}
\end{subequations}
With the use of various functions $f(x)$, $g_1(x)$ and $g_2(x)$, 
we are able to obtain various forms for the deformation 
in the case of the $su(2,1)$-algebra given in the 
framework of three kinds of boson operators. 
We will omit to give their explicit forms. 

As was shown in the above, we performed the deformation 
of the $su(2,1)$-algebra by deforming ${\hat T}_{\pm}^0$ and 
${\hat S}_{\pm}^0$ which are shown in the relation 
(\ref{4-4a}) and (\ref{4-4b}). 
It may be clear from the structure of $\ket{c}$ that these 
are regarded as the raising and the lowering operators for 
the minimum weight state constructed by ${\hat b}_2$-boson. 
In (A), we showed that in order to get the orthogonal sets, 
there exist eight possibilities for choosing the raising 
and the lowering operators together with the boson operators 
constructing the minimum weight state. 
Therefore, our present treatment can be applied to the other 
seven cases and we obtain various deformations. 
In the cases treated in (I) and (II), the deformation was 
uniquely performed.

\section{Discussions}

The aim of this paper is to formulate, in the framework of the 
deformed boson scheme, the time-dependent variational method for 
many-body system constituted of three kinds of boson operators. 
The trial state $\ket{c}$ contains three complex parameters 
for the variation and to perform the variation, we have to calculate 
the expectation value of the Hamiltonian for the state $\ket{c}$. 
As can be seen in the relation (\ref{3-18}) and (\ref{3-20}), 
the expectation value of the Hamiltonian contains the normalization 
constant $\Gamma$ and its derivatives for the parameters. 
Therefore, it may be inevitable to investigate $\Gamma$ in more 
detail than that shown in the relations (\ref{3-11}) and (\ref{3-11-2}). 

For the above-mentioned aim, let us investigate a rather simple 
example shown in the following : 
\begin{equation}\label{5-1}
{\wtilde f}({\hat N}_a)=1 \ , \qquad
{\wtilde g}_1({\hat N}_{b_1})=\sqrt{1+\frac{{\hat N}_{b_1}}{2t}} \ ,
\qquad
{\wtilde g}_2({\hat N}_{b_2})^{-1}=\sqrt{1+{\hat N}_{b_2}} \ .
\end{equation}
As is discussed later, $t$ denotes a sufficiently large constant. 
The relation (\ref{5-1}) gives us the forms 
\begin{subequations}\label{5-2}
\begin{eqnarray}
& &f(n)^2=1 \ , 
\label{5-2a}\\
& &g_1(n)^2=\left(1+\frac{n-1}{2t}\right)\cdot
\left(1+\frac{n-2}{2t}\right)
\cdots \left(1+\frac{1}{2t}\right)\cdot 1 \ , \label{5-2b}
\end{eqnarray}
\end{subequations}
\begin{equation}\label{5-3}
g_2(n)^{-2}=n! \qquad\qquad\qquad\qquad\qquad\qquad\qquad\qquad\qquad
\end{equation}
The form (\ref{5-3}) leads us to 
\begin{equation}\label{5-4}
\sum_{l=0}^\infty \frac{(|\delta|^2)^l}{l!}g_2(l)^{-2}
=\sum_{l=0}^\infty (|\delta|^2)^l
=\frac{1}{1-|\delta|^2} \ . \qquad (|\delta|^2 < 1)
\end{equation}
Then, the normalization constant $\Gamma$ given in the 
relation (\ref{3-11}) is written as 
\begin{equation}\label{5-5}
\Gamma=\Gamma_{\gamma}\cdot \frac{1}{1-|\delta|^2} \ .
\end{equation}
The relations (\ref{3-17a}) and (\ref{3-17b}) give us 
\begin{subequations}\label{5-6}
\begin{eqnarray}
& &u(\partial \Gamma_{\gamma}/\partial u)\cdot \Gamma_{\gamma}^{-1}
=|x_1|^2+|x_2|^2 \ , 
\label{5-6a}\\
& &v(\partial \Gamma_{\gamma}/\partial v)\cdot \Gamma_{\gamma}^{-1}
=|x_2|^2 \ . 
\label{5-6b}
\end{eqnarray}
\end{subequations}
The relation (\ref{3-17c}) leads us to 
\begin{equation}\label{5-7}
|\delta|^2=(|y|^2-|x_2|^2)\cdot (1+|y|^2-|x_2|^2)^{-1} \ .
\end{equation}
Here, $u$ and $v$ are defined in the relation (\ref{2-6c}). 

As was already mentioned, $t$ is sufficiently large. 
Then, we take into account the effects of $t^{-1}$ 
in the linear order : 
\begin{equation}\label{5-8}
g_1(n)^2=1+\frac{1}{4t}n(n-1) \ .
\end{equation}
In the case of $f(n)^2$ and $g_1(n)^2$ shown in the relations 
(\ref{5-2a}) and (\ref{5-8}), respectively, the relations 
(\ref{3-11b}) and (\ref{3-11c}) give us 
\begin{subequations}\label{5-9}
\begin{eqnarray}
& &F_0(0)=1 \ , \qquad F_r(0)=0 \quad \hbox{\rm for}\quad
r=1,2,3,\cdots , 
\label{5-9a}\\
& &G_{10}(0)=1 \ , \qquad G_{12}(0)=\frac{1}{2t} \ , \qquad
G_{1r}(0)=0 \quad \hbox{\rm for}\quad
r=1,3,4,\cdots . \qquad\qquad
\label{5-9b}
\end{eqnarray}
\end{subequations}
Then, $\Gamma_{\gamma}$ given in the relation (\ref{3-11a}) can be 
written in the form 
\begin{subequations}
\begin{eqnarray}
\Gamma_{\gamma}&=&\Gamma^0+\frac{1}{2t}{\wtilde \Gamma} \ , 
\label{5-10}\\
\Gamma^0&=&S(u, -v) \ , 
\label{5-10a}\\
{\wtilde \Gamma}&=&\frac{1}{2}u^2\frac{\partial^2}{\partial u^2}
S(u, -v) \nonumber\\
&=&S(u, -v)\frac{u^2}{(1-u)^4}\left[
(1-u)^2+2(1-u)v+\frac{1}{2}v^2\right] \ .
\label{5-10b}
\end{eqnarray}
\end{subequations}
Here, $S(u, -v)$ is defined in the relation (\ref{2-6b}). 
With the use of the relations (\ref{5-6a}) and (\ref{5-6b}), 
together with the relation (\ref{5-10}), we can determine 
$u$ and $v$, i.e., $|\gamma_1|^2$ and $|\gamma_2|^2$, as 
functions of $|x_1|^2$, $|x_2|^2$ and $|y|^2$. 
In the framework of the linear order for $t^{-1}$, 
the relations (\ref{5-6a}) and (\ref{5-6b}) can be 
written in the following form : 
\begin{subequations}\label{5-11}
\begin{eqnarray}
& &u(\partial \Gamma^0/\partial u)\cdot (\Gamma^0)^{-1}
+\frac{1}{2t}u\frac{\partial}{\partial u}({\wtilde \Gamma}/\Gamma^0) 
=|x_1|^2+|x_2|^2 \ , 
\label{5-11a}\\
& &v(\partial \Gamma^0/\partial v)\cdot (\Gamma^0)^{-1}
+\frac{1}{2t}v\frac{\partial}{\partial v}({\wtilde \Gamma}/\Gamma^0) 
=|x_2|^2 \ . 
\label{5-11b}
\end{eqnarray}
\end{subequations}
In order to solve the relation (\ref{5-11}) in the linear order 
for $1/(2t)$, we decompose $u$ and $v$ as follows : 
\begin{equation}\label{5-12}
u=u^0+\frac{1}{2t}{\wtilde u} \ , \qquad 
v=v^0+\frac{1}{2t}{\wtilde v} \ .
\end{equation}
The parts $u^0$ and $v^0$ are determined by the relation 
\begin{subequations}\label{5-13}
\begin{eqnarray}
& &u^0\left[(\partial \Gamma^0/\partial u)\cdot (\Gamma^0)^{-1}
\right]^0=|x_1|^2+|x_2|^2 \ , 
\label{5-13a}\\
& &v^0\left[(\partial \Gamma^0/\partial v)\cdot (\Gamma^0)^{-1}
\right]^0=|x_2|^2 \ . 
\label{5-13b}
\end{eqnarray}
\end{subequations}
Here, $[Z(u,v)]^0$ denotes the quantity $Z$ at the point 
$u=u^0$ and $v=v^0$. 
With the use of the relation (\ref{5-10a}), $u^0$ and $v^0$ 
are determined in the form 
\begin{eqnarray}\label{5-14}
& &u^0=|x_1|^2\cdot (1+|x_1|^2+|x_2|^2)^{-1} \ , \nonumber\\
& &v^0=|x_2|^2\cdot (1+|x_2|^2)\cdot|x_1|^{-2} \ .
\end{eqnarray}
Then, the zero-th order for $|\gamma_1|^2$ and $|\gamma_2|^2$ 
is given as 
\begin{eqnarray}\label{5-15}
\left[|\gamma_1|^2\right]^0 &=&|x_1|^2\cdot (1+|x_1|^2+|x_2|^2)^{-1} \ , 
\nonumber\\
\left[|\gamma_2|^2\right]^0 &=&|x_2|^2\cdot(1+|x_2|^2)
(1+|x_1|^2+|x_2|^2)^{-1} \nonumber\\
& &\times (1+|y|^2-|x_2|^2)(|y|^2-|x_2|^2)^{-1} \ . 
\end{eqnarray}

Next, we investigate the method to determine ${\wtilde u}$ and 
${\wtilde v}$. The right-hand sides of the relations 
(\ref{5-11a}) and (\ref{5-11b}) are of the zero-th order for 
$1/(2t)$. 
Therefore, the first order terms for $1/(2t)$ on the 
left-hand sides should vanish. From this idea, we obtain 
the following relations : 
\begin{subequations}\label{5-16}
\begin{eqnarray}
& &\Gamma_{uu}{\wtilde u}+\Gamma_{uv}{\wtilde v}=-B_u  \ , 
\label{5-16a}\\
& &\Gamma_{vu}{\wtilde u}+\Gamma_{vv}{\wtilde v}=-B_v  \ . 
\label{5-16b}
\end{eqnarray}
\end{subequations}
Here, $\Gamma_{uu}$, etc. are defined as 
\begin{eqnarray}
& &\Gamma_{uu}=\left[\partial \Gamma^0/\partial u \cdot 
(\Gamma^0)^{-1}+u\frac{\partial}{\partial u}
\left(\partial \Gamma^0/\partial u \cdot (\Gamma^0)^{-1}\right)
\right]^0 \ , 
\nonumber\\
& &\Gamma_{uv}=\left[u\frac{\partial}{\partial v}
\left(\partial \Gamma^0/\partial u \cdot (\Gamma^0)^{-1}\right)
\right]^0 \ , 
\nonumber\\
& &\Gamma_{vu}=\left[v\frac{\partial}{\partial u}
\left(\partial \Gamma^0/\partial v \cdot (\Gamma^0)^{-1}\right)
\right]^0 \ , 
\nonumber\\
& &\Gamma_{vv}=\left[\partial \Gamma^0/\partial v \cdot 
(\Gamma^0)^{-1}+v\frac{\partial}{\partial v}
\left(\partial \Gamma^0/\partial v \cdot (\Gamma^0)^{-1}\right)
\right]^0 \ , 
\label{5-17}\\
& &B_{u}=\left[u\frac{\partial}{\partial u}
\left({\wtilde \Gamma}/\Gamma^0\right)\right]^0 \ , \qquad
B_{v}=\left[v\frac{\partial}{\partial v}
\left({\wtilde \Gamma}/\Gamma^0\right)\right]^0 \ .
\label{5-18}
\end{eqnarray}
More explicitly, the above quantities are expressed as follows : 
\begin{eqnarray}
& &\Gamma_{uu}=
(1-u^0)^{-2}+v^0(1+u^0)(1-u^0)^{-3} \ , \nonumber\\
& &\Gamma_{uv}=u^0(1-u^0)^{-2} \ , \nonumber\\
& &\Gamma_{vu}=v^0(1-u^0)^{-2} \ , \nonumber\\
& &\Gamma_{vv}=u^0(1-u^0)^{-1} \ , 
\label{5-19}\\
& &B_u=(u^0)^2\left[
2(1-u^0)^2+2(2+u^0)(1-u^0)v^0+(1+u^0)(v^0)^2\right]
(1-u^0)^{-5} \ , \nonumber\\
& &B_v=v^0(u^0)^2\left[2(1-u^0)+v^0\right](1-u^0)^{-4} \ .
\label{5-20}
\end{eqnarray}
With the help of the relation (\ref{5-16}), together with 
the relations (\ref{5-19}) and (\ref{5-20}), we can 
determine ${\wtilde u}$ and ${\wtilde v}$ as functions of 
$|x_1|^2$, $|x_2|^2$ and $|y|^2$. 

As was promised in \S 3, we investigate the form 
$\partial \Gamma/\partial |\gamma_2|^2\cdot \Gamma^{-1}$. 
This quantity can be expressed as 
\begin{eqnarray}\label{5-21}
\partial \Gamma/\partial |\gamma_2|^2 \cdot \Gamma^{-1}
&=&\frac{|\delta|^2}{|\gamma_1|^2}\cdot\frac{1}{v}\cdot v\left(
\partial \Gamma_{\gamma}/\partial v\right)\cdot \Gamma_\gamma^{-1} 
\nonumber\\
&=&|x_2|^2(|y|^2-|x_2|^2)(1+|y|^2-|x_2|^2)^{-1}(uv)^{-1} \ .
\end{eqnarray}
The part $(uv)^{-1}$ can be approximated in the form 
\begin{equation}\label{5-22}
(uv)^{-1}=(u^0v^0)^{-1}
\left[1-\frac{1}{2t}(u^0v^0)^{-1}(v^0{\wtilde u}
+u^0{\wtilde v})\right] \ .
\end{equation}
The solutions obtained in the above lead us to the 
following results : 
\begin{eqnarray}
& &|x_2|^2(|y|^2-|x_2|^2)(1+|y|^2-|x_2|^2)^{-1}(u^0v^0)^{-1}
\nonumber\\
& &\quad =(1+|x_1|^2+|x_2|^2)(1+|x_2|^2)^{-1}
(|y|^2-|x_2|^2)(1+|y|^2-|x_2|^2)^{-1} \ , \qquad
\label{5-23}\\
& &-(u^0v^0)^{-1}(v^0{\wtilde u}+u^0{\wtilde v})\nonumber\\
& &\quad 
=(|x_1|^2+|x_2|^2)(1+|x_1|^2+|x_2|^2)(1+|x_2|^2)^{-1} \ . 
\label{5-24}
\end{eqnarray}
Then, we have 
\begin{eqnarray}\label{5-25}
\partial \Gamma/\partial |\gamma_2|^2 \cdot \Gamma^{-1}
&=&(1+|x_1|^2+|x_2|^2)(1+|x_2|^2)^{-1}\nonumber\\
& &\times (|y|^2-|x_2|^2)(1+|y|^2-|x_2|^2)^{-1}\nonumber\\
& &\times \left[1+\frac{1}{2t}(|x_1|^2+|x_2|^2)
(1+|x_1|^2+|x_2|^2)(1+|x_2|^2)^{-1}\right] .\qquad\quad
\end{eqnarray}
With the aid of the form (\ref{5-25}), we can calculate the 
expectation value of ${\hat S}_+$ defined in the relation 
(\ref{4-4b}) for $\ket{c}$, the explicit form of which 
is given in the relation (\ref{3-20}) : 
\begin{eqnarray}\label{5-26}
\bra{c}{\hat S}_+\ket{c}
&=&\bra{c}{\hat b}_1^*\sqrt{1+\frac{{\hat N}_{b_1}}{2t}}
\left(\sqrt{1+{\hat N}_{b_2}}\right)^{-1}{\hat b}_2 \ket{c}
\nonumber\\
&=&x_2^*\sqrt{1+|x_1|^2(1+|x_2|^2)^{-1}} \nonumber\\
& &\times \frac{1}{\sqrt{2t}}\sqrt{
2(t+|x_1|^2)+|x_2|^2-|x_1|^2(1-|x_1|^2)(1+|x_2|^2)^{-1}}
\nonumber\\
& &\times \sqrt{(|y|^2-|x_2|^2)(1+|y|^2-|x_2|^2)^{-1}} \ .
\end{eqnarray}

In (B), we have described a possible description of 
time-evolution of statistically mixed state for the 
following Hamiltonian expressed in terms of the present 
notations : 
\begin{subequations}\label{5-27all}
\begin{eqnarray}
& &{\hat H}={\hat K}_{b_1}+{\hat K}_{b_2}+{\hat V}_{b_1b_2} \ , 
\label{5-27}\\
& &{\hat K}_{b_1}=\omega {\hat N}_{b_1} \ , 
\label{5-27a}\\
& &{\hat K}_{b_2}=\omega {\hat N}_{b_2} \ , 
\label{5-27b}\\
& &{\hat V}_{b_1b_2}=-\gamma\sqrt{2t}\cdot i({\hat S}_+-{\hat S}_-) 
\ . 
\label{5-27c}
\end{eqnarray}
\end{subequations}
Here, $\omega$ and $\gamma$ are positive constants. Under the above 
Hamiltonian, we described the statistically mixed state of the 
system composed of the boson operator $({\hat b}_1, {\hat b}_1^*)$. 
In this case, the system composed of the boson operator 
$({\hat b}_2, {\hat b}_2^*)$ plays a role of the external 
environment. Further, we introduced the boson operator 
$({\hat a}, {\hat a}^*)$ in order to describe the mixed state 
in terms of the phase space doubling. We reinvestigate roughly the 
Hamiltonian (\ref{5-27all}) in the present scheme. 
The expectation value of ${\hat H}$ for $\ket{c}$, $H$, is 
given in the following form : 
\begin{subequations}\label{5-28all}
\begin{eqnarray}
& &H=2\omega(k-1)+\omega\tau-\omega t-\gamma\cdot i
(\tau_+^0-\tau_-^0)\cdot \rho \sigma \ , 
\label{5-28}\\
& &k=|y|^2/2+1 \ , \qquad (k=\bra{c}{\hat K}\ket{c}) 
\label{5-28a}\\
& &\tau=t+|x_1|^2 \ , \qquad (|x_1|^2=\bra{c}{\hat N}_a\ket{c})
\label{5-28b}\\
& &\tau_+^0=x_2^*\sqrt{2\tau+|x_2|^2} \ , \qquad
\tau_-^0=x_2\sqrt{2\tau+|x_2|^2} \ , 
\label{5-28c}\\
& &\rho=\sqrt{1+\frac{|x_1|^2}{1+|x_2|^2}}\sqrt{
1-\frac{|x_1|^2}{1+|x_2|^2}\cdot\frac{1-|x_1|^2}{2\tau+|x_2|^2}} \ , 
\label{5-28d}\\
& &\sigma=\sqrt{\frac{2(k-1)-|x_2|^2}{1+2(k-1)-|x_2|^2}} \ .
\label{5-28e}
\end{eqnarray}
\end{subequations}
The above is obtained by rewriting $\bra{c}{\hat S}_+\ket{c}$ 
shown in the relation (\ref{5-26}). We can see that the Hamiltonian 
(\ref{5-28}) is reduced to $H_{su(2,1)}$ in the relation 
(\ref{5-5}) in (B) under the following approximation : 
\begin{equation}\label{5-29}
\rho \approx 1 \ , \qquad \sigma \approx 1 \ .
\end{equation}
Of course, $t$, i.e., $\tau$ should be large. 
In (B), we investigated the time-evolution of the statistically 
mixed state for the Hamiltonian (\ref{5-28}) under the 
condition (\ref{5-29}) and various results could be obtained. 
The condition $\rho\approx 1$ tells us that $|x_2|^2$ is 
larger than $|x_1|^2$. Since 
$|x_1|^2=\bra{c}{\hat N}_a \ket{c}$ and 
$|x_1|^2+|x_2|^2=\bra{c}{\hat N}_{b_1}\ket{c}$, the amplitude of 
the motion induced by the boson $({\hat b}_1, {\hat b}_1^*)$ 
is larger than the measure of the statistical mixture. 
The condition $\sigma\approx 1$ implies, as was mentioned in 
(B), 
that the motion induced by the boson $({\hat b}_1, {\hat b}_1^*)$ 
is of sufficiently long periodic motion.

Finally, we give some short comments. In (III), we presented the 
deformed boson scheme for the $su(2,1)$-algebra in three kinds of boson 
operators. Further, we showed that the description of the 
statistically mixed state given in (B) can be reproduced 
completely in the present scheme. In (I) and (II), rather well-known 
points are discussed systematically. But, Part (III) contains 
various aspects of the deformed boson scheme which are not so 
well known as those in (I) and (II). 
In this sense, it may be expected to investigate various 
problems in many-boson systems in the present scheme.


\section*{Acknowledgements}

This work was completed when two of the authors (Y. T. \& M. Y.) stayed 
at Universidade de Coimbra in the middle of August 2001. 
They wish to thank Professor J. da Provid\^encia, one of 
the co-authors, for his warm hospitality. 
One of the authors (Y. T.) was partially supported by a Grant-in-Aid 
for Scientific Research from the Ministry of Education, Culture, Sports, 
Science and Technology (No.13740159).

\appendix
\section{A method for calculating the normalization constant 
for the state $\ket{c}$}

For the preparation of giving a method for calculating the 
normalization constant, we present two mathematical formulae. 
Let $R(n)$ denote a function for $n$ ($=0, 1, 2, \cdots)$, 
which obeys $|R(n)| < \infty$. The function $R(n)$ 
can be expressed in the following form : 
\begin{equation}\label{A-1}
R(n)=\sum_{r=0}^n \frac{R_r(0)}{r!} n_r \ .
\end{equation}
Here, $R_r(0)$ and $n_r$ are defined as 
\begin{eqnarray}
& &R_r(0)=\sum_{s=0}^{r} \frac{r! (-)^{r-s}}{s!(r-s)!}
R(s) \ , 
\label{A-2}\\
& &n_r=\cases{ 1 \ , \qquad\qquad\qquad\qquad\qquad (r=0) \cr
               n(n-1)\cdots(n-r+1) \ . \ \ (r=1, 2, 3, \cdots)}
\label{A-3}
\end{eqnarray}
The proof is easy. If noting the relation $(1-1)^{n-s}=1$ for 
$n=s$ and 0 for $n\neq s$, the straightforward calculation of the 
right-hand side of the relation (\ref{A-1}) for the forms 
(\ref{A-2}) and (\ref{A-3}) leads us to the left-hand side 
of the relation (\ref{A-1}). 
The meaning of $R_r(0)$ can be given in the following way : 
Let us define the function
\begin{equation}\label{A-4}
R_r(k)=\sum_{s=0}^{r}\frac{r!(-)^{r-s}}{s!(r-s)!}R(s+k) \ . 
\qquad (k=0, 1, 2, \cdots)
\end{equation}
The case $k=0$ is nothing but the form (\ref{A-2}). 
For $R_r(k)$, we find the relation 
\begin{equation}\label{A-5}
\frac{R_{r-1}(k+1)-R_{r-1}(k)}{(k+1)-k}=R_r(k) \ .
\end{equation}
The relation (\ref{A-5}) tells us that $R_r(k)$ is the 
difference of $r$-th order at the point $k$. 
Therefore, $R_r(0)$ means the difference of $r$-th 
order at the point $k=0$. 
The above is the meaning of $R_r(0)$. The quantity $n_r$ 
defined in the relation (\ref{A-3}) can be extended to 
the case $r=n+1,\ n+2, \cdots$ : 
\begin{equation}\label{A-6}
n_r=n(n-1)\cdots (n-n+1)(n-(n+1)+1)\cdots (n-r+1)=0 \ .
\end{equation}
Therefore, the form (\ref{A-1}) can be expressed as 
\begin{equation}\label{A-7}
R(n)=\sum_{r=0}^\infty \frac{R_r(0)}{r!} n_r \ .
\end{equation}

Our next task is to prove the following relation : 
\begin{equation}\label{A-8}
\frac{(m+n)!}{m!}=\sum_{r=0}^{n}\frac{(n!)^2(-)^{n-r}}{(r!)^2(n-r)!} 
\frac{(m+n+r)!}{(m+n)!} \ . \qquad (m,\ n=0,\ 1,\ 2, \cdots)
\end{equation}
For the case $n=0$, the relation (\ref{A-8}) holds and, 
then, the cases $n=1, 2, 3, \cdots$ are interesting. 
For this aim, we introduce two polynomials of degree $n$
$(n=1,2,3,\cdots)$ for the real variable $x$ : 
\begin{eqnarray}
& &P_n(x)=(x+1)(x+2)\cdots(x+n) \ , 
\label{A-9}\\
& &Q_n(x)=(-)^{n} n! 
+\sum_{r=1}^n \frac{(n!)^2(-)^{n-r}}{(r!)^2(n-r)!}
(x+n+1)(x+n+2)\cdots(x+n+r) \ .\qquad\quad
\end{eqnarray}
We calculate the values of $P_n(x)$ and $Q_n(x)$ at $(n+1)$ 
points of $x$ given as 
\begin{equation}\label{A-11}
x_m=-(m+n+1) \ . \qquad 
(m=0, 1, 2, \cdots , n)
\end{equation}
The case of $P_n(x)$ is as follows : 
\begin{equation}\label{A-12}
P_n(x_m)=(-)^n \frac{(m+n)!}{m!} \ .
\end{equation}
On the other hand, we have 
\begin{eqnarray}\label{A-13}
Q_n(x_m)&=&
\sum_{r=0}^m (-)^{n-r}\frac{(n!)^2}{(r!)^2(n-r)!}(-)^r 
\frac{m!}{(m-r)!} \nonumber\\
&=&
(-)^n n!\sum_{r=0}^m \frac{n!}{r!(n-r)!} 
\frac{m!}{r!(m-r)!} \nonumber\\
&=&
(-)^n n!\cdot \frac{(m+n)!}{m!n!} \nonumber\\
&=&(-)^n \frac{(m+n)!}{m!} \ .
\end{eqnarray}
Therefore, the following relation is obtained : 
\begin{equation}\label{A-14}
P_n(x_m)=Q_n(x_m)  \ . \qquad (m=0, 1, 2, \cdots , n)
\end{equation}
The above means that the values of the two polynomials 
of degree $n$, 
$P_n(x)$ and $Q_n(x)$, are equal to each other at 
the $(n+1)$ different points of $x$. 
Then, we have 
\begin{equation}\label{A-15}
P_n(x)=Q_n(x) \ .
\end{equation}
Therefore, the relation (\ref{A-15}) tells us that at $x=m$ 
($m=0, 1, 2, \cdots$), we get 
\begin{subequations}\label{A-16s}
\begin{eqnarray}
& &P_n(m)=Q_n(m) \ , 
\label{A-16}\\
P_n(m)&=&(m+1)(m+2)\cdots (m+n) =\frac{(m+n)!}{m!} \ , 
\label{A-16a}\\
Q_n(m)&=&(-)^n n!+\sum_{r=1}^n \frac{(n!)^2(-)^{n-r}}{(r!)^2(n-r)!}
(m+n+1)(m+n+2)\cdots (m+n+r) \nonumber\\
&=&\sum_{r=0}^n \frac{(n!)^2(-)^{n-r}}{(r!)^2(n-r)!}\cdot
\frac{(m+n+r)!}{(m+n)!} \ .
\label{A-16b}
\end{eqnarray}
\end{subequations}
Thus, we could prove the relation (\ref{A-8}). 

With the aid of the relations (\ref{A-7}) (or (\ref{A-1})) 
and (\ref{A-8}), we calculate the normalization constant 
$\Gamma_\gamma$ for the state 
\begin{equation}\label{A-17}
\ket{\gamma}= \left(\sqrt{\Gamma_\gamma}\right)^{-1}
\exp\left(\xi {\hat c}^*{\wtilde \phi}({\hat N}_c)
\cdot {\hat d}^* {\wtilde \psi}({\hat N}_d)\right)
\exp\left(\eta {\hat d}^*{\wtilde \psi}({\hat N}_d)\right) \ket{0} \ .
\end{equation}
Here, $\xi$ and $\eta$ denote complex parameters. 
The operators $({\hat c} , {\hat c}^*)$ and $({\hat d} , {\hat d}^*)$ 
are boson operators and ${\hat N}_c$ and ${\hat N}_d$ are 
boson number operators $({\hat N}_c={\hat c}^*{\hat c}$ ; 
${\hat N}_d={\hat d}^*{\hat d})$. 
The operators ${\wtilde \phi}({\hat N}_c)$ and 
${\wtilde \psi}({\hat N}_d)$ denote functions of ${\hat N}_c$ and ${\hat N}_d$ 
with the properties suitable for our discussion in this paper. 
The state $\ket{\gamma}$ is rewritten in the form 
\begin{eqnarray}\label{A-18}
\ket{\gamma}
&=&\left(\sqrt{\Gamma_\gamma}\right)^{-1}
\exp \left((\xi{\hat c}^*{\wtilde \phi}({\hat N}_c)+\eta)\cdot
{\hat d}^*{\wtilde \psi}({\hat N}_d)\right)\ket{0} \nonumber\\
&=&\left(\sqrt{\Gamma_\gamma}\right)^{-1}
\sum_{n=0}^\infty \frac{\psi(n)}{n!}(\xi {\hat c}^*
{\wtilde \phi}({\hat N}_c)+\eta)^n ({\hat d}^*)^n \ket{0} \ .
\end{eqnarray}
Here, $\psi(n)$ is defined as 
\begin{equation}\label{A-19}
\psi(n)=\cases{1 \ , \qquad\qquad\qquad\qquad\qquad\qquad (n=0) \cr
               {\wtilde \psi}(n-1){\wtilde \psi}(n-2)\cdots
               {\wtilde \psi}(0) \ . \quad (n=1, 2, 3, \cdots)}
\end{equation}
Further, $\ket{\gamma}$ is rewritten as 
\begin{equation}\label{A-20}
\ket{\gamma}=\left(\sqrt{\Gamma_\gamma}\right)^{-1}
\sum_{n=0}^\infty \frac{\psi(n)}{n!}\sum_{m=0}^n 
\frac{n!}{m!(n-m)!}\xi^m \eta^{n-m}\phi(m) 
({\hat c}^*)^m({\hat d}^*)^n \ket{0} \ .
\end{equation}
Here, $\phi(m)$ is defined as 
\begin{equation}\label{A-21}
\phi(m)=\cases{1 \ , \qquad\qquad\qquad\qquad\qquad\qquad (m=0) \cr
               {\wtilde \phi}(m-1){\wtilde \phi}(m-2)\cdots
               {\wtilde \phi}(0) \ . \quad (m=1, 2, 3, \cdots)}
\end{equation}
Then, $\Gamma_\gamma$ is expressed in the form 
\begin{subequations}\label{A-22s}
\begin{eqnarray}
& &\Gamma_\gamma = \sum_{n=0}^{\infty}
\sum_{m=0}^n \Psi(n)\Phi(m)\frac{n!}{m![(n-m)!]^2}u^n v^{n-m} \ .
\label{A-22}\\
& &\Psi(n)=\psi(n)^2 \ , \qquad \Phi(m)=\phi(m)^2 \ .
\label{A-22a}
\end{eqnarray}
\end{subequations}
The quantities $u$ and $v$ are related to $|\xi|^2$ and $|\eta|^2$ 
in the form 
\begin{equation}\label{A-23}
|\xi|^2=u \ , \qquad |\eta|^2=uv \ .
\end{equation}
We apply the relation (\ref{A-7}) (or (\ref{A-1})) to the form 
(\ref{A-22a}) : 
\begin{equation}\label{A-24}
\Phi(m)=\sum_{r=0}^\infty \frac{\Phi_r(0)}{r!} m_r 
=\sum_{r=0}^m \Phi_r(0)\frac{m!}{r!(m-r)!} \ .
\end{equation}
Then, $\Gamma_\gamma$ given in Eq.(\ref{A-22}) is expressed as 
\begin{equation}\label{A-25}
\Gamma_\gamma=\sum_{n=0}^\infty \frac{\Phi_n(0)}{n!}u^n 
\sum_{m=0}^\infty u^m L_m(-v)\Psi(m+n)\frac{(m+n)!}{m!} \ .
\end{equation}
The symbol $L_m(-v)$ denotes the Laguerre polynomial. 
Further, we apply the relation (\ref{A-7}) to 
$\Psi(m+n)$ in Eq.(\ref{A-25}) : 
\begin{equation}\label{A-26}
\Psi(m+n)=\sum_{r=0}^\infty \frac{\Psi_r(0)}{r!}(m+n)_r \ .
\end{equation}
Then, $\Gamma_\gamma$ is, further, rewritten as 
\begin{equation}\label{A-27}
\Gamma_\gamma=\sum_{n=0}^\infty \sum_{m=0}^\infty \sum_{r=0}^\infty 
\frac{\Phi_n(0)}{n!}\frac{\Psi_r(0)}{r!}u^{n+m}
L_m(-v)(m+n)_r \frac{(m+n)!}{m!} \ .
\end{equation}
With the aid of the relation (\ref{A-8}), we have 
\begin{eqnarray}\label{A-28}
(m+n)_r\frac{(m+n)!}{m!}
&=& (m+n)_r\sum_{s=0}^n \frac{(n!)^2(-)^{n-s}}{(s!)^2(n-s)!}\cdot
\frac{(m+n+s)!}{(m+n)!} \nonumber\\
&=&\sum_{s=0}^n \frac{(n!)^2(-)^{n-s}}{(s!)^2(n-s)!}\cdot
\frac{(m+n+s)!}{(m+n-r)!} \nonumber\\
&=&\sum_{s=0}^n \frac{(n!)^2(-)^{n-s}}{(s!)^2(n-s)!}
(m\!+\!n\!+\!s)(m\!+\!n\!+\!s\!-\!1)\cdots (m\!+\!n\!-\!r\!+\!1) 
\ .\nonumber\\
\end{eqnarray}
The relation (\ref{A-28}) gives us 
\begin{eqnarray}\label{A-29}
\Gamma_\gamma&=&
\sum_{n=0}^\infty \sum_{m=0}^\infty \sum_{r=0}^\infty \sum_{s=0}^n 
\frac{\Phi_n(0)}{n!}\frac{\Psi_r(0)}{r!}(-)^{n-s}
\frac{(n!)^2}{(s!)^2(n-s)!} \nonumber\\
& &\times (m\!+\!n\!+\!s)(m\!+\!n\!+\!s\!-\!1)
\cdots (m\!+\!n\!-\!r\!+\!1)u^{m+n}L_m(-v) \ .
\end{eqnarray}
We note the relation 
\begin{equation}\label{A-30}
(m+n+s)(m+n+s-1)\cdots(m+n-r+1)u^{m+n}
=u^r \left[\left(\frac{\partial}{\partial u}\right)^{r+s}
u^{m+n+s}\right] \ .
\end{equation}
Thus, we have 
\begin{eqnarray}\label{A-31}
\Gamma_\gamma&=&\sum_{n=0}^\infty \sum_{r=0}^\infty \sum_{s=0}^n 
\frac{\Phi_n(0)}{n!}\frac{\Psi_r(0)}{r!} (-)^{n-s}
\frac{(n!)^2}{(s!)^2(n-s)!} \nonumber\\
& &\qquad\qquad \times 
u^r \left(\frac{\partial}{\partial u}\right)^{r+s}
u^{n+s}\left(\sum_{m=0}^\infty u^m L_m(-v)\right) \nonumber\\
&=&
\sum_{r=0}^\infty \frac{\Psi_r(0)}{r!}u^r \sum_{n=0}^\infty 
\sum_{s=0}^n (-)^{n-s}\frac{\Phi_n(0)n!}{(s!)^2(n-s)!} 
\nonumber\\
& &\qquad\qquad \times 
\left(\frac{\partial}{\partial u}\right)^{r+s}
\left[ u^{n+s}S(u, -v)\right] \ .
\end{eqnarray}
Here, we used the well-known formula for the generating 
function of the Laguerre polynomial : 
\begin{equation}\label{A-32}
\sum_{n=0}^\infty L_n(x)t^n 
=S(t, x) \ , \qquad
S(t, x)=\frac{e^{-x\frac{t}{1-t}}}{1-t} \ .
\end{equation}

We have another form for $\Gamma_\gamma$ by applying the 
relation (\ref{A-7}) to the following case : 
\begin{equation}\label{A-33}
\Psi(m+n)(m+n)!=\sum_{r=0}^\infty \frac{\Xi_r(0)}{r!}(m+n)_r \ .
\end{equation}
Of course, $\Xi_r(0)$ denotes the difference with $r$-th 
degree for $\Psi(m+n)(m+n)!$. With the use of the relation 
(\ref{A-33}), $\Gamma_\gamma$ given in the relation (\ref{A-25}) 
can be rewritten as 
\begin{equation}\label{A-34}
\Gamma_\gamma=\sum_{n=0}^\infty \sum_{m=0}^\infty \sum_{r=0}^\infty 
\frac{\Phi_n(0)\Xi_r(0)}{n!m!r!}(m\!+\!n)\cdots(m\!+\!n\!-\!r\!+\!1)
u^{m+n}L_m(-v) \ .
\end{equation}
We note the relation 
\begin{equation}\label{A-35}
(m+n)(m+n-1)\cdots(m+n-r+1)u^{m+n}
=u^r \left(\frac{\partial}{\partial u}\right)^r u^{m+n} \ .
\end{equation}
Then, $\Gamma_\gamma$ is expressed as 
\begin{eqnarray}\label{A-36}
\Gamma_\gamma&=&
\sum_{n=0}^\infty \sum_{r=0}^\infty
\frac{\Phi_n(0)}{n!}\frac{\Xi_r(0)}{r!}u^r
\left(\frac{\partial}{\partial u}\right)^r u^n 
\left(\sum_{m=0}^\infty \frac{u^m}{m!}L_m(-v)\right) \nonumber\\
&=&\sum_{r=0}^\infty \frac{\Xi_r(0)}{r!}u^r 
\left(\frac{\partial}{\partial u}\right)^r 
\left[e^u I_0(2\sqrt{uv})\sum_{n=0}^\infty \frac{\Phi_n(0)}{n!}
u^n\right] \ .
\end{eqnarray}
Here, $I_0(2\sqrt{uv})$ denotes the modified Bessel function 
with the 0-th degree : 
\begin{equation}\label{A-37}
I_0(2\sqrt{uv})=J_0(2i\sqrt{uv})
=\sum_{n=0}^\infty \frac{(uv)^n}{(n!)^2} \ .
\end{equation}

It should be noted that the notations used in this Appendix 
correspond to the following : 
\begin{eqnarray}\label{A-38}
& &\xi \longrightarrow \gamma_1 \ , \qquad
\eta \longrightarrow \gamma_2 \delta \ , \nonumber\\
& &u \longrightarrow |\gamma_1|^2 \ , \qquad 
v \longrightarrow |\gamma_1|^{-2}|\gamma_2|^2|\delta|^2  \ , 
\nonumber\\
& &({\hat c} , {\hat c}^*) \longrightarrow ({\hat a} , {\hat a}^*) \ , 
\qquad
({\hat d} , {\hat d}^*) \longrightarrow ({\hat b}_1 , {\hat b}_1^*) \ , 
\nonumber\\
& &
{\wtilde \phi}({\hat N}_c) \longrightarrow 
{\wtilde f}({\hat N}_a) \ , \qquad
{\wtilde \psi}({\hat N}_d) \longrightarrow 
{\wtilde g}_1({\hat N}_{b_1}) \ , \nonumber\\
& &\Phi(n) \longrightarrow 
F(n) \ , \qquad
\Psi(n) \longrightarrow 
G_1(n) \ .
\end{eqnarray}

\end{document}